# Analytical Evaluation of Unfairness Problem in Wireless LANs

Ahmed Mohamedou, and Mohamed Othman

**Abstract**—The number of users using wireless Local Area Network is increasing exponentially and their behavior is changing day after day. Nowadays, users of wireless LAN are using huge amount of bandwidth because of the explosive growth of some services and applications such as video sharing. This situation imposes massive pressure on the wireless LAN performance especially in term of fairness among wireless stations. The limited resources are not distributed fairly in saturated conditions. The most important resource is the access point buffer space. This importance is a result of access point being the bottleneck between two different types of networks. These two types are wired network with relatively huge bandwidth and wireless network with much smaller bandwidth. Also the unfairness problem is keep getting worse because of the greedy nature Transmission Control Protocol (TCP). In this paper, we conduct a comprehensive study on wireless LAN dynamics and proposed a new mathematical model that describes the performance and effects of its behavior. We validate the proposed model by using the simulation technique. The proposed model was able to produce very good approximation in most of the cases. It also gave us a great insight into the effective variables in the wireless LAN behavior and what are the dimensions of the unfairness problem.

**Index Terms**— Access Point, Buffer Size, Congestion Window, TCP, Wireless LAN.

——————————— ◆ ———————————

## 1 INTRODUCTION

Today, Wireless networks are becoming very essential part in any industry or service. It is hard to see a public place such as airports, hospitals, coffee shops, universities and shopping centers without wireless local area network that serve their customers as a part of attracting more customers. More and more people are using wireless LAN as preferred LAN in their home or office networks.

This huge demand makes research communities to give more concentration on these wireless LAN's. It is obvious that wireless LAN's are still relatively new technology that has many problems. Problems such as limited bandwidth (compared to wired LAN's), high error rate and unfairness among different stations. The last mentioned problem differs from other problems that it is not an engineering problem but it is a result of mechanisms and protocols that are used in upper layers. These mechanism and protocols were written and designed in seventies of twentieth century by people who did not have in their minds the wireless concept. And when wireless technology came up, these mechanisms and protocols were already wide spread and it was really hard to change everything for wireless sake. As a result, the only way to solve problems such as unfairness is to study some parts of these mechanisms and protocols and try to enhance the performance of these parts which leads to overall performance improvement. Keep in mind that improving one aspect of these mechanisms and protocols may degrade other different aspects. For example, some ideas that already proposed to improve the unfairness problem reduce the network utilization.

If each one of these mechanism and protocols is analyzed separately, it will produce in most case perfect results. However, some problems such as the unfairness only arise when these mechanisms are mixed. The unfairness problem is getting more importance every day since wireless LANs are becoming larger in term of covered area and bandwidth. It has been acknowledged in literature extensively, [1,2,3,7].

Qian et al [1] have conducted comprehensive experiments using both of test bed and simulation approaches to show the effect of 802.11 MAC protocol and cross-layer interaction on TCP fairness in wireless LANs. They proposed two queue management techniques to ease the unfairness among TCP flows.

Eun-Chan et al [2] differentiate between TCP-induced asymmetry and MAC-induced asymmetry and show how these two kinds of asymmetry amplify the unfairness problem. They introduced a cross-layer feedback technique to guarantee the fairness in wireless LANs.

Our work in this paper is motivated by Pilosof et al [3]. They studied the unfairness among upstream and downstream wireless stations. They presented a comprehensive simulation study and a mathematical model which is the core of our work.

The rest of paper is organized as follow. Section two analyze the unfairness problem and explains it sources. Section three develops the analytical model and state number of assumptions. Section four represents the simulation experiments and discusses its results. Section five concludes the research.

————————————————

- *Ahmed Mohamedou is with Department of Communications Technology and Networks, Universiti Putra Malaysia, 43400 UPM Serdang, Selangor D.E., Malaysia.*
- *Mohamed Othman is with Department of Communications Technology and Networks, Universiti Putra Malaysia, 43400 UPM Serdang, Selangor D.E., Malaysia.*

## 2 PROBLEM OVERVIEW

In wireless LAN, the most used architecture is Infrastructure where there are number of wireless stations (STA) communicating with wired stations (STA') through Access Point (AP). These wireless stations are divided into two kinds, sending stations (UP) and receiving stations (DOWN).

Sending stations send data to their corresponding wired stations through the access point. While receiving station wait for the data coming from their corresponding wired stations to be sent by the access point. The unfairness problem can be understood from two observations:

- Only sending stations and the access point compete to occupy the wireless channel while receiving stations just keep listening to the channel and let the access point to compete on behave of them. This kind of behavior makes sending stations to dictate the wireless channel and do not let receiving stations to have a fair share of the channel because they are not competing to occupy the channel.

- The second observation is related to TCP. The access point here has to move traffic between two different networks, the wired network which has huge bandwidth and the wireless network which has much smaller bandwidth. This means that for sure the outgoing buffer of the access point on the wireless LAN side will become full very fast and incoming packets from corresponding wired stations will be dropped. These packets are either data packets or acknowledgment packets of TCP protocol in transport layer. Data packets are coming to receiving stations and when one of these packets is dropped, the Congestion Control Mechanism in corresponding wired station will be invoked which leads to reduce its sending rate and as result the accessing channel share of the receiving station will be reduced. While acknowledgment packets are coming to sending stations in the wireless LAN and when one of these packets is dropped nothing will change if the next acknowledgment packet is delivered before Retransmission-Time Out (RTO) of dropped packet elapsed. This behavior will make the accessing channel share of sending stations increase while the accessing channel share of receiving station decrease.

To evaluate the problem severity, number of simulation experiments were conducted using NS2 simulator [4]. The performance matrices that have been used are Throughput Ratio and Jain Fairness Index. Since we want to evaluate the performance of wireless LAN against Access Point buffer size, we used the later as the simulation variable in all experiments. The simulation is repeated many times with different number of wireless stations.

### 2.1 One UP station and one DOWN station

In figure 1, we can see that as the buffer size increases, the throughput ratio approach to one. This keeps happening till the access point buffer size is equal or greater than 84, which is the summation of TCP window size of both stations (TCP window size maximum value is set to 42). At this point, the throughput ration will be equal one because there are sufficient resources (in term of access point buffer size) to serve both of the wireless stations.

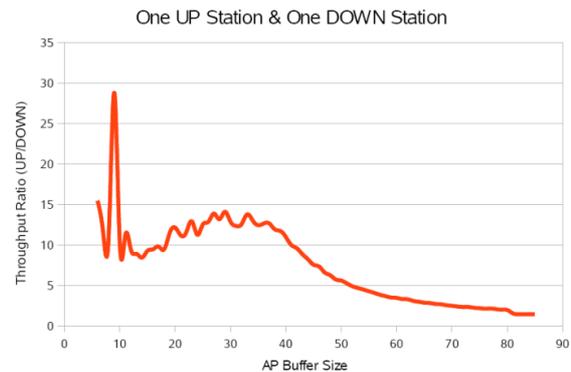

Fig. 1. Throughput Ratio of One UP station and One DOWN Station.

### 2.2 Two UP stations and two DOWN stations

From figures 1 and 2, we describe that when the number of wireless stations increases, the problem is getting worse and worse. Keep in mind; the number of UP stations and DWON stations in both figures has the same ratio. However, in figure 2 the highest throughput ratio value is 320 while in figure 1 the highest throughput value is 28.

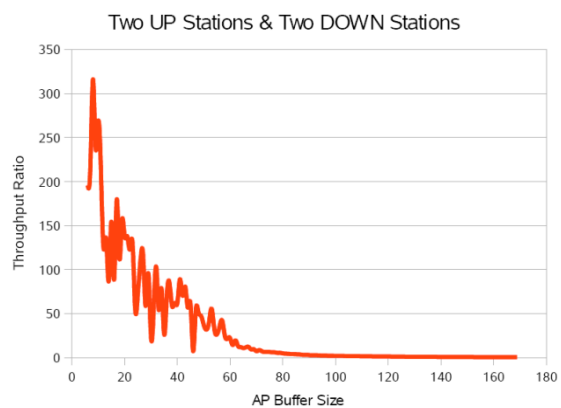

Fig. 2. Throughput Ratio of Two UP stations and two DOWN Stations.

### 2.3 Two UP stations and one DOWN station

Figure 3 shows the fact that when number of UP stations is more than the number of DOWN stations, the severity of problem is getting much higher than when both of UP stations and DOWN stations have the same population. In this figure, the highest throughput ratio is 510 and as the ratio of UP stations number to DOWN stations number increases, the throughput ratio will increase as well.

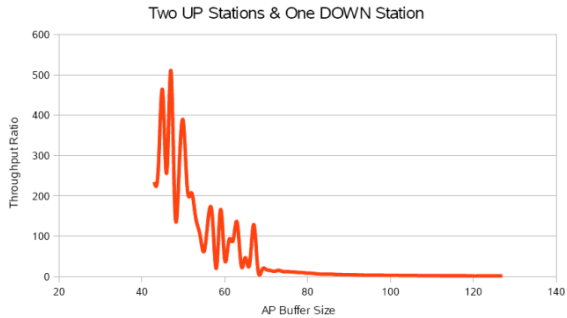

Fig. 3. Throughput Ratio of Two UP stations and one DOWN station.

### 2.4 One UP station and two DOWN stations

Figure 4 represents the simulation experiment when the ratio of UP stations number to DOWN stations number is less than one, which means there are more DOWN stations than UP stations. Despite the later fact, UP stations still able to get more share of wireless channel than the DOWN stations. UP stations keep doing that till the access point buffer size become equal to 84, which is equal to one UP station TCP maximum window size multiplied by two. This due to fact that there are enough resources to serve two stations, one UP station and one DOWN station, and any extra share that is given to the third station (which is DOWN station) will make throughput ratio approach to zero in favor of DOWN stations.

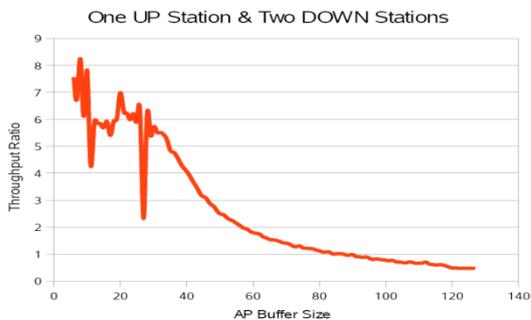

Fig. 4. Throughput Ratio of one UP station and two DOWN stations.

### 2.5 Fairness Index

Figure 5 shows that the performance of wireless LAN in term of fairness in all case has the same behavior and pattern. It starts between 0.4 and 0.6 and keeps increasing as the access point buffer size increased till it reach its maximum value of one where there are sufficient resources to serve all wireless stations in the wireless LAN.

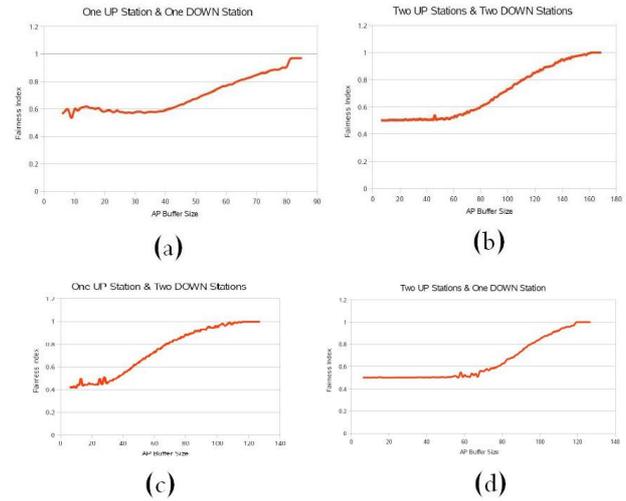

Fig. 5. Fairness index, (a) one UP station and one DOWN station. (b) two UP stations and two DOWN stations. (c) one UP station and two DOWN stations. (d) two UP stations and one DOWN station.station.

## 3 MATHEMATICAL MODEL

In this section, a mathematical model that describe the wireless LAN's performance will be proposed and developed. This mathematical model is based on number of variables such as access point buffer size, TCP maximum congestion window size, number of DOWN stations and number of UP stations. It derives to approximate the wireless LAN performance in term of throughput ratio. The queue model M/M/1/B and TCP model will be combined together to produce the proposed mathematical model.

### 3.1 Assumptions

As it well known, in the beginning of any analytical study we need to state number of assumptions that will helps to make developing the mathematical is feasible. These assumptions should not be broad so that the mathematical model under construction becomes trivial. The simplest way of improving any analytical study is by omitting one or more assumptions. In this paper we used the following assumptions:

1. No specific version of TCP is considered in particular. We considered the general dynamic of TCP protocol.
2. In TCP, for every data packet that is delivered there is one acknowledgment packet generated.
3. Round Trip Times (RTTs) are the same for all wireless stations in the wireless LAN.
4. Access point buffer waiting delay for queued packets coming to the wireless LAN is neglectable.
5. All UP stations are sending their data packets with rate of *w/RTT* which is the maximum sending rate of any TCP connection.
6. All TCP connections are using the same maximum window size.

## 3.2 Queue Modeling

Let's consider wireless LAN as queue system of model M/M/1/B. As explained by Kleinrock [5] in his famous book Queuing Systems, every queue system has a ratio, which is defined by:

$$\rho = \frac{Arrival\ rate}{Serving\ rate} \quad (1)$$

the probability of this system having the queue size of B is calculated as follow:

$$\Pr(queue\ size = B) = \frac{(1-\rho)}{(1-\rho^{B+1})}\rho^B \quad (2)$$

To ease the use of equation (2), we should rewrite it so that the exponent part is in on side of the equation and the non-exponent is in the other side. It goes as follow:

$$\rho^B = \frac{Pr}{1 + \rho\,(Pr - 1)} \quad (3)$$

Now we need another equation that approximate $Pr$ and this is what next section is about.

## 3.3 TCP Modeling

In this section, we are going to use a model that developed by Pdhye in [6] to approximate TCP receiving rate of DOWN stations.

$$R = \frac{1}{RTT}\sqrt{\frac{3}{2\,Pr}} \quad (4)$$

Note that for modeling the sending rate of UP stations, we are using assumption 5.

## 3.4 Proposed Model

Since we are trying to analyze the throughput ratio of the wireless LAN against its access point buffer size, we need first to define the Throughput Ratio ($R$). Before that we need to state notations that are going to be used. They are:

TABLE 1
MODELING NOTATIONS

| Notation | Description |
|---|---|
| $R_U$ | Sending rate of one uplink station |
| $R_D$ | Receiving rate of one downlink station |
| $U$ | Number uplink stations |
| $D$ | Number downlink stations |
| $B$ | Access point buffer size |
| $w$ | TCP maximum window size |
| $R$ | Ratio between the total sending rates of uplink stations and the total receiving rate of downlink stations which is equal to : $$R = \frac{DR_D}{UR_U} \quad (5)$$ |

In the base work [7], authors considered only one DOWN station and one UP station in the wireless LAN ($R = R_D/R_U$) with the ability to increase DOWN stations with simple extra modification. However, any attempt to increase the number of UP stations in their model will results in complex modifications and bad approximations to the wireless LAN performance. Here, we improve their model by making it expandable to any number of UP stations and DOWN stations.

Based on equation (1), we need to define the arrival rate and the serving rate. The arrival rate here is equal to:

$$Arrival\ rate = UR_U + DR_D \quad (6)$$

where $DR_D$ is representing the data packets coming to the DOWN stations in the wireless LAN and $UR_U$ is representing the acknowledgement packets coming to the UP stations in the wireless LAN.

The serving rate is equal to:

$$Serving\ rate = R_U \quad (7)$$

Since access point in DCF is treated as same as one of UP stations in the wireless LAN. Therefore, by substituting equations (7), (6), (5) and (1), we will have:

$$\rho = U(1 + R) \quad (8)$$

Now equation (3) can be written as follow:

$$[U\,(1+R)]^B = \frac{Pr}{1 - U\,(1+R)(Pr-1)} \quad (9)$$

To get rid of $Pr$ in equation (9), we need to find another way to compute it. One way is to calculate $R$ by using TCP sending rate model which is developed in the previous section. We need to do simple modification in equation (4) as follow:

$$R_D = \frac{1}{RTT}\left(\sqrt{\frac{3}{2Pr}} + E\right) \quad (10)$$

where $E$ represents the chance of a DOWN station getting served by the extra space in the access point buffer when there is more space than what all UP stations need. $E$ can be evaluated as follow:

$$E = \frac{3(B - Uw)}{4D}\ ,where\ (B - Uw) > 0$$
$$E = 0\ ,Otherwise \quad (11)$$

Now, by using assumption 5 and equations (10) and (5), we will have:

$$Pr = \frac{3D^2}{2(UwR - DE)^2} \quad (12)$$

By substituting equation (12) in equation (9), then equation (9) can be written as follow:

$$(1 + R)^B[c_3 R^3 + c_2 R^2 + c_1 R + c_0] = \frac{3D^2}{U^B} \quad (13)$$

where

$$c_3 = -2U^3 w^2 \quad (14)$$

$$c_2 = 4DEU^2w - 2U^3w^2 + 2U^2w^2 \quad (15)$$

$$c_1 = 3UD^2 - 2UD^2E^2 + 4DEU^2w - 4DEUw \quad (16)$$

$$c_0 = 3UD^2 - 2UD^2E^2 + 2D^2E^2 \quad (17)$$

Now we need to find a way to get rid of the exponent in $(1 + R)^B$. In the base work [7], authors assumed that $(1 + R)^B$ is approximately equal to *(1+BR)* if $R$ is small. But this will not be valid as the size of the wireless LAN increases which is our motivation in this paper.

Another way to simplify equation (13) is to use logarithm as follow:

$$\ln((1+R)^B[c_3R^3 + c_2R^2 + c_1R + c_0])$$
$$= \ln\left(\frac{3D^2}{U^B}\right) \quad (18)$$

By using logarithm properties, we have

$$B\ln(1+R) + \ln(c_3R^3 + c_2R^2 + c_1R + c_0)$$
$$= \ln\left(\frac{3D^2}{U^B}\right) \quad (19)$$

In the equation (19), since we have a constant on the right side of the equation, we can take the derivative of the equation to get rid of the logarithm:

$$\frac{B}{1+R} + \frac{3c_3R^2 + 2c_2R + c_1}{c_3R^3 + c_2R^2 + c_1R + c_0} = 0 \quad (20)$$

After simple algebraic operations

$$(Bc_3 + 3c_3)R^3 +$$
$$(3c_3 + 2c_2 + Bc_2)R^2 + \quad (21)$$
$$(2c_2 + c_1 + Bc_1)R + c_1 + Bc_0 = 0$$

This is third degree equation. Finding its zero will provide us with the throughput ratio of the wireless LAN. There are many ways to its zero. One way is to use the Cubic Formula [8]. It is long formula that has been used since fourteenth century to solve third degree equations. Or we can use routines provided in NAG Fortran Library [9] or MATPACK C++ Library [10]. We chose to rewrite our own solving utility in C# .based on the cubic formula [8].

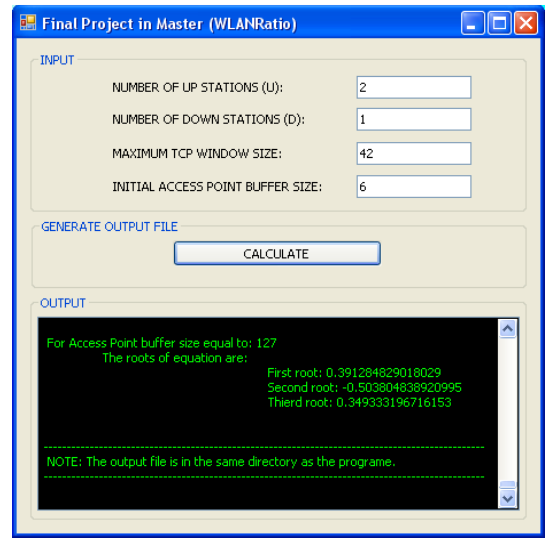

Fig. 6. Interface of our computing utility.

## 4 RESULTS AND DISCUSSION

In this research, numbers of simulation experiments were conducted to validate the proposed model. NS2 [4] was used as the simulator. Simple topology is used to construct the network as shown in the figure 7.

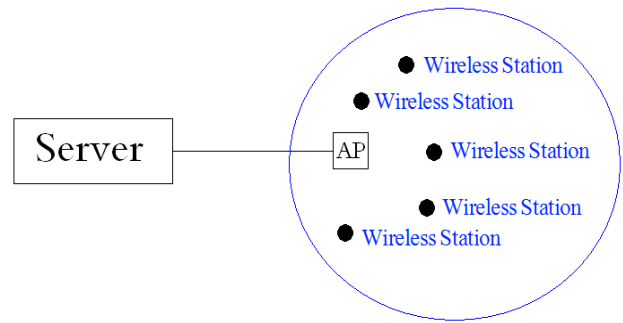

Fig. 7. Simulation Topology.

Here, we have number of wireless stations communicating with one wired server through the access point. For all TCP agents, the maximum congestion window size is set to 42 packets. We used IEEE 802.11b with maximum bandwidth of 11Mbps. The RTS/CTS mechanism was switched off. The wired link between the access point and the server has bandwidth of 100 Mbps. In all queues, we used first in first out queue management technique. All queue sizes were set to 1000 packets except for the access point buffer so that no buffer overflow will happen in any queue except the access point buffer. We used FTP agent as the application in all simulation experiments. We ran each experiment for 100 seconds. Table 2 summarizes the setup parameter:

TABLE 2
SIMULATION PARAMETERS

| Parameter | Value |
| --- | --- |
| TCP window size | 42 |
| Wireless bandwidth | 11 Mbps |
| RTS/CTS mechanism | OFF |
| Wired bandwidth | 100 Mbps |
| Queue management | FIFO |
| Queue size | 1000 |
| Application | FTP |
| Simulation time | 100 seconds |

We ran different experiments for different scenarios. These scenarios represent the wireless LAN with different populations as follow:
- Scenario 1: Simulation of one UP station and one DOWN station.
- Scenario 2: Simulation of two UP stations and two DOWN stations.
- Scenario 3: Simulation of one UP station and two DOWN stations.
- Scenario 4: Simulation of two UP stations and one DOWN station.

We also compared the proposed model with the model in base work [7]. Using only algebraic operations to extend the base work model will result a fourth degree equation as follow:

$$[-2x^3w^2B]R^4 + [B(2x^2w^2 - 2x^3w^2 + 4x^2wyE) - 2x^3w^2]R^3 \\ + [B(4x^2wyE - 2xy^2E^2 + 3xy^2 - 4xwyE) \\ + (2x^2w^2 - 2x^3w^2 + 4x^2wyE)]R^2 \\ + [B(3xy^2 - 2xy^2E^2 + 2y^2E^2) \\ + (4x^2wyE - 2xy^2E^2 + 3xy^2 - 4xwyE)]R \\ + (3xy^2 - 2xy^2E^2 + 2y^2E^2) \\ - 3y^2/x^B = 0 \qquad (22)$$

We called the proposed model in next figures as *New Analytical Model* and the algebraic extended base work model as *Old Analytical Model*.

### 4.1 Scenario 1

First, we ran the same simulation experiment that has been conducted in the base work [7]. We needed to validate that the proposed model perform as same as the base model in the simple scenario of one UP station and one DOWN station.

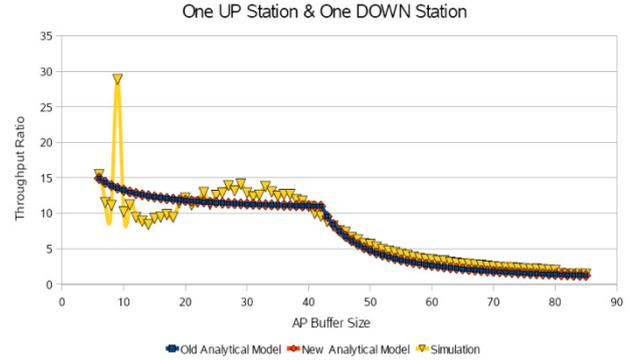

Fig. 8. Comparison of one UP station and one DOWN station scenario.

Figure 8 represents the comparison between the simulation and the proposed models in this work and the base work. We can see that the models are giving very good approximations especially when the access point buffer size is more than 42 which is the maximum TCP window size. The reason behind that is when there is enough space in access point buffer to serve the sending rate of the UP station; any extra space in the access point buffer will be designated to serve the DOWN station receiving rate and there will be no competition between the UP and DOWN stations. This will lead to smooth performance of the wireless LAN.

### 4.2 Scenario 2

In this experiment, we kept the ratio of UP stations number to the DOWN stations number equal to one which means both of UP stations and DOWN stations have the same population size.

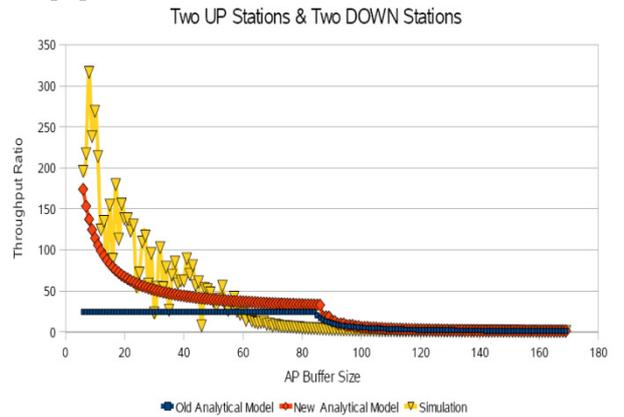

Fig. 9. Comparison of two UP stations and two DOWN stations scenario.

In figure 9, we have two regions, the first one is from zero to 84 and the second one is from 84 to 168. In the first region the simulation results are not stable but the proposed model manage to give good approximation to the overall behavior. In the second region the simulation results are smoother and as the access point buffer size ap-proach to 168, the proposed model achieves perfect match. Note that when the limit of access point buffer size goes to 84 from both sides, the model approximations do not fit the simulation results. This maybe due to the fact that DOWN stations are getting served more than what the model assume since UP stations do not need 100% of access point resources in that region to achieve their maximum sending rate. Regarding the base work model, we can see that it fails terribly to describe wireless LAN behavior and our proposed model is very good enhancement.

### 4.3 Scenario 3

In this experiment, we let the ratio of UP stations number to the DOWN stations number be less than one which means there are more DOWN stations than UP stations in the wireless LAN.

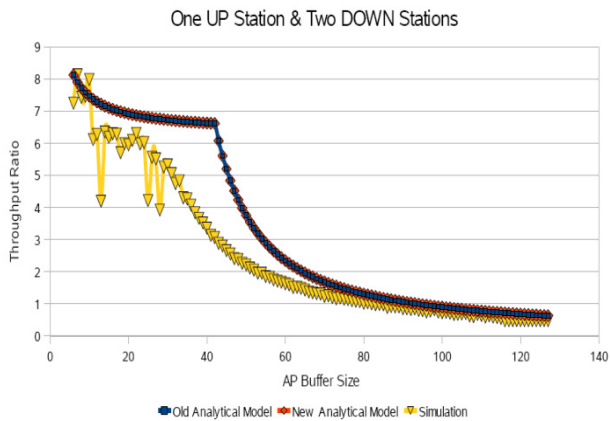

Fig. 10. Comparison of one UP station and two DOWN stations scenario.

In figure 10, we also have two regions, the first one is from zero to 42 and the second one is from 42 to 126. In the first region the simulation results are not stable but the proposed model manage to give good approximation to the overall behavior. In the second region the simulation results are smoother and as the access point buffer size approach to 126, the proposed model achieves perfect match. Note that when the limit of access point buffer size goes to 42 from both sides, the model approximations do not fit the simulation results. As we mentioned before this maybe due to the fact that DOWN stations are getting served more than what the model assume since UP stations do not need 100% of access point resources in that region to achieve their maximum sending rate. The perfect match between the new model and the old one is due to the fact that we are using only one UP station. If UP stations population increases, the weaknesses of old model will be more obvious.

### 4.4 Scenario 4

In this experiment, we have more UP stations than DOWN stations. Therefore, the ratio of UP stations number to the DOWN stations number is more than one.

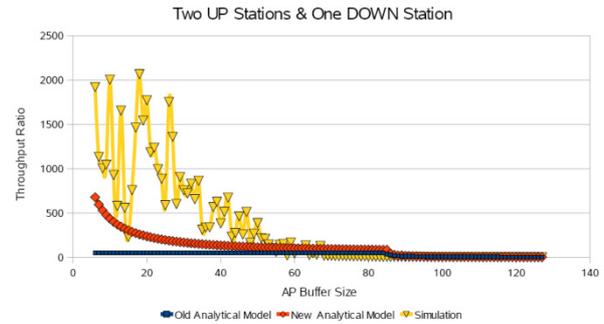

Fig. 11. Comparison of two UP stations and one DOWN station scenario.

In this scenario, the severity of problem is increased since the ratio of UP stations number to the DOWN stations number is more than one. However, the proposed model managed to give good approximations for the most of simulation either in term of general behavior when the simulation results are not stable or in term of perfect matching when the simulation results are stable. For the region when the limit of access point buffer size goes to 84 from both sides, the model approximations do not fit the simulation results. And as we mentioned before this maybe due to the fact that DOWN station is getting served more than what the model assume since UP stations do not need 100% of access point resources in this region to achieve their maximum sending rate. We also can see from figure 11 that the old model is almost useless in such a scenario.

## 5 Conclusion

Wireless Local Area Networks are increasing in popularity day after day. They are becoming the first choice for any computer network. As their bandwidth and covering area are increasing, the numbers of served users are increasing as well. This will lead to unfairness problem since the behavior of these users is classified into two types. The first type is when the user is send data most of the time. While the second type is when the user is receiving data most of the time.

In this paper, the unfairness problem was studied and received a comprehensive investigation. Then, we introduced designing and implementing the proposed mathematical model. At the beginning, we stated the assumptions that have been used in developing the proposed model. Then we went through the modification of M/M/1 queue system so that it suite for our purpose. After that we went through step by step illustration of proposed model development. Then we provided information about the simulation experiments that have been conducted to validate the proposed model. Next we went through each simulation experiment and gave detailed discussion.

At the end we reached a conclusion that in general the proposed model performs very well and give excellent approximations especially when the ratio of UP stations number to the DOWN stations number is approaching to zero which means there are more DOWN stations than UP stations. This case is the most common case in the deployed wireless LANs nowadays. There are always more DOWN stations than UP stations. The last fact makes this model very useful to study wireless LANs.

For the future work, we are planning to extend the mathematical model to work in the large wireless networks especially fourth generation mobile networks candidate such as WiMax and LTE-Advance. In these kinds of networks there are two uses of the wireless stations. Either they are used to access the internet, so the wireless stations in these networks are the same as the one in wireless LANs. Or they are used as phone devices. In the later case, the wireless station will become an UP station and DOWN station at the same time because the amount of data that has been sent is the same as the amount of data that has been received since it is used for voice conversation.

Another idea is to apply the proposed model in IEEE 802.11n. In this standard the access point is not the network bottleneck since both of the wired and wireless networks have equivalent bandwidth. However, packets in the access point buffer will sever larger delay since the access point is serving network with large bandwidth.

## REFERENCES

[1] Wu Q., Gong M., Williamson C., "TCP fairness issues in IEEE 802.11 wireless LANs ," Computer Communications, 31 (10), pp. 2150-2161, 2008.

[2] Eun-Chan Park, Dong-Young Kim, Hwangnam Kim, Chong-Ho Choi, "A Cross-Layer Approach for Per-Station Fairness in TCP over WLANs," IEEE Transactions on Mobile Computing, vol. 7, no. 7, pp. 898-911, July 2008.

[3] S. Pilosof, R. Ramjee, D. Raz, Y. Shavitt, P. Sinha, "Understanding TCP fairness over wireless LAN," in IEEE INFOCOM, 2003, pp. 863-872.

[4] NS2, The Network Simulator ns-2, Available: http://www.isi.edu/nsnam/ns/

[5] L. Kleinrock, Queuing Systems, Volume 1: THEORY. John Wiley and Sons, 1973.

[6] J. Padhye, V. Firoiu, D. Towsley, and J. Kurose, "Modeling TCP Throughput: a Simple Model and its Empirical Validation," in ACM Sigcomm, 1998, pp. 303-314.

[7] S. Pilosof, R. Ramjee, Y. Shavitt, and P. Sinha, "Understanding TCP fairness over wireless LAN," in IEEE INFOCOM '05, 2005, p. 863–872.

[8] Weisstein, Eric W. "Cubic Formula." From MathWorld--A Wolfram Web Resource. http://mathworld.wolfram.com/CubicFormula.html

[9] NAG Fortran Library [Online]. Available: http://www.nag.com/numeric/FL/FLdescription.asp

[10] Matpack C++ Numerics and Graphics Library [Online]. Available:http://www.matpack.de/